\documentclass[12pt,epsfig]{article}
\usepackage[dvips]{graphicx}
\usepackage{amsfonts}
\usepackage{amsmath}
\usepackage{amssymb}
\newtheorem{postulate}{Postulate}[section]
\newcommand{\bpost}{\begin{postulate} \rm}
\newcommand{\epost}{\end{postulate}}
\newtheorem{theorem}{Theorem}[section]
\newcommand{\btheo}{\begin{theorem} \it}
\newcommand{\etheo}{\end{theorem}}
\newtheorem{proposition}{Proposition}[section]
\newcommand{\bprop}{\begin{proposition}\it}
\newcommand{\eprop}{\end{proposition}}
\newtheorem{corollary}{Corollary}[section]
\newcommand{\bcorol}{\begin{corollary} \it}
\newcommand{\ecorol}{\end{corollary}}
\newtheorem{lemma}{Lemma}[section]
\newcommand{\blem}{\begin{lemma}\it}
\newcommand{\elem}{\end{lemma}}
\newtheorem{remark}{Remark}[section]
\def\brem{\begin{remark} \rm }
\def\erem{\end{remark}\ \\}
\newtheorem{example}{Example}[section]
\def\bexam{\begin{example} ~~\\ \rm}
\def\eexam{\end{example}}
\newtheorem{definition}{Definition}[section]
\newcommand{\bdefi}{\begin{definition} \rm }
\newcommand{\edefi}{\end{definition}\ \\}

\begin{document}


\title{\textbf{Models for Equilibrium BEC}\\
\textbf{Superradiance}}\maketitle

\begin{center}

\setcounter{footnote}{0}
\renewcommand{\thefootnote}{\arabic{footnote}}

\vspace{0.2 cm}

\setcounter{footnote}{0}
\renewcommand{\thefootnote}{\arabic{footnote}}

{\bf Joseph V. Pul\'{e}\footnote{ Email : Joe.pule@ucd.ie},
Andr\'{e} Verbeure \footnote{ Email : andre.verbeure@fys.kuleuven.ac.be},  \\ and \\
Valentin A. Zagrebnov\footnote{ Email :
zagrebnov@cpt.univ-mrs.fr}}

\vspace{0.5 cm}

\textbf{Department of Mathematical Physics, \\ University College,
Belfield, Dublin 4, Ireland}{$^{\textbf{1}}$}\\

\vspace{0.2 cm}

\textbf{Instituut voor Theoretische Fysika, \\ Katholieke
Universiteit Leuven, Celestijnenlaan 200D,\\ 3001 Leuven, Belgium
}{$^{\textbf{2}}$}\\

\vspace{0.2 cm}

\textbf{Universit\'e de la M\'editerran\'ee and Centre de Physique
Th\'eorique,\\ CNRS-Luminy-Case 907,
 13288 Marseille, Cedex 09, France}{$^{\textbf{3}}$}

\vspace{1 cm}

{\bf Abstract}

\end{center}
Motivated by recent experiments with superradiant Bose-Einstein
Condensate (BEC) we consider simple microscopic models describing
rigorously the interference of the two cooperative phenomena, BEC
and radiation, in thermodynamic equilibrium. Our resuts in
equilibrium confirm the presence of the observed superradiant
light scattering from BEC: (a) the equilibrium superradiance
exists only below a certain transition temperature; (b) there is
superradiance and matter-wave (BEC) enhancement due
to the coherent coupling between light and matter. 
\vspace{0.5 cm}\\
\noindent\textbf{ Running title} : BEC and
superradiance
\bigskip
\bigskip

\renewcommand{\thefootnote}{\fnsymbol{footnote}}


\setcounter{footnote}{0} \renewcommand{\thefootnote}{\arabic{footnote}} %

\newpage

\noindent \textbf{1.} This letter is motivated by the interest in
Bose-Einstein Condensation (BEC) of bosons in \textit{traps} and
in particular, by the recent discovery of the Dicke superradiance
and BEC matter-wave \textit{amplification}, see e.g. \cite {K},
\cite {KI-1}, \cite {KI-2}. In all these experiments, the
condensate was illuminated with a $Q$-mode laser beam (so-called
\lq\lq dressing beam") and then the BEC atoms scatter photons from
this beam into another mode and receive the corresponding recoil
momentum producing a coherent \lq\lq four-wave mixing" of light
and atoms \cite {KI-2}. Notice that Girardeau in 1978, \cite{Gi},
had already anticipated the possible existence of these phenomena.

The irradiation of a Bose-Einstein condensate can be considered as
an external action on it. The effect of other external agents have
been studied, for instance the influence of boundary conditions
(\cite{R} and \cite{VVZ}) and that of an external field, scaled
with the volume so as to retain the space homogeneity  \cite{P}.
In the first case it is well known that attractive boundary
conditions enhance condensation. Here condensation occurs even in
one dimension and is a consequence of a discrete point in the
spectrum. In the other case one also finds an enhancement of
condensation compared to that of the free Bose gas depending on
the behaviour of the potential.

In the present letter we consider \textit{two} simple models which
show that some particular interactions with a one-mode radiation can
\textit{enhance} the conventional BEC of the perfect Bose gas
(PBG). This can be interpreted as a coherent coexistence of the
BEC and the \textit{condensation} of photons, which is a type of
\textit{equilibrium} Dicke superradiance induced by BEC, see
e.g. \cite {HL}, \cite {BZT}, \cite {FSV}.

We consider a system of non-interacting bosons (PBG) of
mass $m$ enclosed in a cubic box $\Lambda \subset \Bbb{R}^{d}$ of
volume $V=\left| \Lambda \right| =L^{d}$ centered at the origin
with periodic boundary conditions. The Hamiltonian for the PBG is
\begin{equation}\label{PBG}
T_{\Lambda }:=\sum_{k\in \Lambda ^{*}}\epsilon
_{k}a_{k}^{*}a_{k}
\end{equation}
where
\[
k\in\Lambda ^{*}=\left\{ k\in \Bbb{R}^{d}:\text{ }k=\frac{2\pi n}{L}\text{, }n\in \Bbb{Z}^d\right\},
\]
$\epsilon _{k}=\hbar^{2}k^{2}/2m $ is the kinetic energy of one particle and
$a_{k}^{*}$ and $a_{k}$ are the usual boson creation and annihilation operators
corresponding to momentum
$\hbar k $, satisfying the commutation relations
\[
[a_k,a_{k'}^*]=\delta_{k k'}, \ \ \ [a_k,a_{k'}]=[a_k^*,a_{k'}^*]=0.
\]
In both of our models the PBG (\ref{PBG}) interacts with a one-mode
\textit{photon} field with Hamiltonian
\begin{equation}\label{photons}
\Omega\, b^{*} b,
\end{equation}
where $b^{*}$ and $b$ are the photon creation and annihilation
operators, satisfying $[b,b^*]=1$, and $\Omega
> 0$ is the energy of a single photon.

In both our models we assume that the photons interact with
bosons \textit{linearly}. The interaction in the first model has
the form:
\begin{equation}\label{int-linear-1}
U_{1 \Lambda}= \frac{1}{2} g_1 \,( a_{Q}^{*} b + a_{Q} b^* ),
\end{equation}
whereas in the second one it is:
\begin{equation}\label{int-linear-2}
U_{2 \Lambda}= \frac{1}{2} g_2 \,( a_{Q'}^{*} b^{*} + a_{Q'} b ).
\end{equation}
Hence the Hamiltonians of our models have the form :
\begin{equation}\label{models 1-2}
H_{1,2 \,\Lambda}= {\tilde T}_\Lambda + \Omega\, b^{*} b + U_{1,2
\,\Lambda},
\end{equation}
where ${\tilde T}_\Lambda:=\sum_{k\in \Lambda,\,k\neq 0}\epsilon _{k}a_{k}^{*}a_{k}$ and $g_{1,2} \geq 0$.

The different forms of interactions (\ref{int-linear-1}) and
(\ref{int-linear-2}) come from two possible mechanisms of the
light-bosons couplings.

It is known that in the cubic box the \textit{conventional} BEC of
the PBG occurs only in the mode $k=0$, see e.g. \cite{vdBLP}.
Since the scattering of the $Q$-mode light is most important on
the macroscopic amount of condensed particles, following \cite{Gi}
and \cite{MM} we retain in the interaction only terms representing
excitation from, and de-excitation back, to the BEC (in conformity with the minimal coupling in electrodynamics):
\begin{equation}\label{int-1}
\widehat{U}_{1 \Lambda}= \frac{1}{2} \frac{\lambda_1
}{\sqrt{V}}\,( a_{Q}^{*} a_{0} b_{Q}  + a_{0}^* a_{Q} b^*_{Q} ),
\end{equation}
with a coupling $\lambda_1 \geq 0$. Now assuming (as in \cite {KI-2})
that the depletion of the condensate can be neglected, one can
simplify (\ref{int-1}) by substituting
\begin{equation}\label{c-numb}
\frac{1}{\sqrt{V}}\,\, a_{0}\,\rightarrow\, \sqrt{\rho_0}\, e^{i
\varphi}\,,\,\frac{1}{\sqrt{V}}\,\, a_{0}^*\,\rightarrow\,
\sqrt{\rho_0}\, e^{- i \varphi},
\end{equation}
where $\rho_0$ is the $k=0$ mode condensate density. This leads to the
interaction (\ref{int-linear-1}), with the coupling constant $g_1 = \lambda_1
\sqrt{\rho_0}$, after a gauge transformation eliminating the zero
mode condensate phase $\varphi$ and after putting $b_{Q} = b$.

The interaction (\ref{int-linear-2}) has as its origin in the
\lq\lq four-wave mixing" mechanism , see e.g. \cite {K}-\cite {KI-2}. A
condensate illuminated by a $Q$-mode \lq\lq dressing" laser beam
(\lq\lq dressed condensate") can spontaneously emit pairs of photons and
recoiling atoms. The simplest way to take this into account is
described by the Hamiltonian \cite {KI-2}:
\begin{equation}\label{int-2}
\widehat{U}_{2 \Lambda}= \frac{1}{2} \frac{\lambda_2
}{\sqrt{V}}\,( a_{Q'}^{*} b_{Q''}^{*} a_{0} b_{Q}  + a_{0}^*
b^*_{Q} a_{Q'} b_{Q''} ),
\end{equation}
where $b_{Q''}^{*}, b_{Q''}$ correspond to superradiated photons
with a wave-vector $Q''= Q - Q'$. If as in  \cite {KI-2} one
neglects the depletion of the $k=0$ mode \lq\lq dressed" condensate and
the $Q$-mode \lq\lq dressing" laser beam, then as in (\ref{c-numb})
the substitution of the corresponding operators by $c$-numbers
gives instead of (\ref{int-2})
the interaction (\ref{int-linear-2}). Now $g_2$ is proportional
to $\lambda_2 \sqrt{\rho_0}$ and the amplitude of the \lq\lq dressing"
laser beam, and $b_{Q''} = b$.

The aim of the present letter is to study the thermodynamic
\textit{equilibrium} properties of the models (\ref{models 1-2})
and the possible \textit{phase transitions} due to the coherent
interaction of \textit{recoiled} condensate atoms with scattered
(\textit{superradiated}) photons.
\\
\\
\noindent \textbf{2.} These models can be solved \textit{exactly}
by canonical transformations diagonalizing
Hamiltonians (\ref{models 1-2}). To establish the thermodynamic
properties we consider the \textit{grand-canonical ensemble}
Hamiltonians
\begin{equation}\label{mu-Ham}
H_{1,2 \,\Lambda}(\mu)= H_{1,2 \,\Lambda} - \mu {\tilde N}_\Lambda,
\end{equation}
where $\mu$ is the
chemical potential and
\begin{equation}\label{part-numb}
{\tilde N}_\Lambda= \sum_{k\in \Lambda,\,k\neq 0}a_{k}^{*}a_{k}\,,
\end{equation}
is the particle number operator for the system excluding the ground state condensate. Since we shall follow a
procedure analogous to that used for the PBG it is useful to
recall some facts about the latter.

For the PBG the finite-volume grand-canonical thermodynamic
functions exist only for $\mu < 0$. Denoting the
\textit{grand-canonical Gibbs state} corresponding to the
Hamiltonian $T_\Lambda$ at inverse temperature $\beta$ by
$\left\langle - \right\rangle_{T_\Lambda}(\mu)$, we have for
$\mu < 0$,
\begin{equation}\label{PBGdens}
\lim_{V\to \infty}\left\langle\frac{N_\Lambda}{V}
\right\rangle_{T_\Lambda}\hskip -0.4 cm(\mu)=\rho_0(\mu)
\end{equation}
where $N_\Lambda={\tilde N}_\Lambda+a^*_0a_0$ and
\begin{equation}\label{PBG-part-dens1}
\rho_0( \mu) = \frac{1}{(2\pi)^d}\int_{\mathbb{R}^d}
 \frac{d^{d}k}{e^{\beta (\epsilon _{k} - \mu)} - 1}.
\end{equation}
For $d\geq 3$, $\rho_{0 c}\equiv\rho_0( 0)<\infty $
and therefore, to be able to access densities higher than $\rho_{0 c}$ in
this case, one has to work with
a given density rather than the chemical potential. We fix the finite volume chemical potential
$\mu_V$ by the equation
\begin{equation}\label{part-dens1}
\rho=\left\langle\frac{N_\Lambda}{V}
\right\rangle_{T_\Lambda}\hskip -0.4 cm( \mu_V).
\end{equation}
One then finds that for $\rho \leq \rho_{0 c}$, $\mu_V \to \mu_\infty$
where $\mu_\infty< 0$ is the unique solution of the equation
\begin{equation}
\rho = \rho_0( \mu).
\end{equation}
On the other hand for $\rho > \rho_{0 c}$,
\begin{equation}
\mu_V = -\frac{1}{\beta V(\rho -\rho_{0 c})}+{\rm O}(1/V^2).
\end{equation}
For the expectation of the zero mode occupation particle density we have
\begin{equation}\label{zero modePBG}
\lim_{V\rightarrow\infty}\left\langle\frac{a_0^*a_0}{V}
\right\rangle_{T_\Lambda}\hskip -0.4cm ( \mu_V) =
\begin{cases}
0 & {\rm for}\ \rho \leq \rho_{0 c},\\
\rho - \rho_{0 c} & {\rm for}\ \rho > \rho_{0 c}.
\end{cases}
\end{equation}

\noindent \textbf{2.1} Returning to our first model
$H_{1\,\Lambda}$ let us now define new boson operators $\xi_1^* ,
\xi_1$ and $\eta_1^* , \eta_1$ by the canonical transformation:
\begin{eqnarray}\label{can-trans}
&& \xi_1 = \cos\vartheta \,\,a_{Q} + \sin\vartheta \,\, b \\
&& \eta_1 = \sin\vartheta \,\,a_{Q} - \cos\vartheta \,\, b
\nonumber \,.
\end{eqnarray}
where $\vartheta$ satisfies the equation
\begin{equation}\label{equ-theta}
    \tan 2\vartheta = - \,\frac{g_1}{\Omega - \epsilon_Q + \mu}.
\end{equation}
In terms of these operators we can write the Hamiltonian
$H_{1\,\Lambda}(\mu)$ in the form
\begin{equation}\label{mu-Ham1-diag}
H_{1\,\Lambda}(\mu) = \sum_{k\in \Lambda ^{*},\,k\neq 0,Q}(\epsilon _{k}- \mu) a_{k}^{*}a_{k} + E_{1\,+} (\mu
, \Omega) \xi_{1}^* \xi_1 + E_{1\,-} (\mu , \Omega) \eta_{1}^*
\eta_1 \,\,,
\end{equation}
where
\begin{eqnarray}\label{E1,2}
&& E_{1\,+} (\mu , \Omega) = \frac{1}{2}\left\{\Omega +
\epsilon_Q - \mu + \sqrt{(\Omega + \mu - \epsilon_Q)^2 +
g_{1}^2}\right\} \\
&& E_{1\,-} (\mu , \Omega) = \frac{1}{2}\left\{\Omega +
\epsilon_Q - \mu - \sqrt{(\Omega + \mu - \epsilon_Q)^2 +
g_{1}^2}\right\} \nonumber \,\,.
\end{eqnarray}
Since $E_{1\,+} (\mu , \Omega)\geq E_{1\,-} (\mu , \Omega)$, the
thermodynamic stability of the Hamiltonian $H_{1\,\Lambda}(\mu)$
requires that
\begin{equation}\label{stab}
E_{1\,-} (\mu , \Omega) \geq 0 \,\,\,\, \,\,\mbox{or} \,\,\,\, \mu
\leq  \mu _c (g_{1},\Omega)\,\,
\end{equation}
where
\begin{equation}\label{stab}
\mu_c(g,\Omega)\equiv \min \left\{0\,,\, -\frac{g_{1}^2}{4\Omega} +
\epsilon_Q\right\}.
\end{equation}
Denoting the \textit{grand-canonical Gibbs state} corresponding to
the Hamiltonian $H_{1 \Lambda}$ by $\left\langle -\right\rangle_{H_{1 \Lambda}}( \mu)$, we have for $\mu < \mu _c
$,
\begin{equation}\label{zero modeV}
\left\langle\frac{a_{Q}^{*}a_{Q}}{V}
\right\rangle_{H_{1 \Lambda}}\hskip -0.4 cm ( \mu)
=\frac{1}{2V}\left\{\left(\frac{1}{e^{\beta E_{1\,+}} - 1} +
\frac{1}{e^{\beta E_{1\,-}} - 1}\right) -\frac{\Omega+\mu -\epsilon_Q}{E_{1\,+} - E_{1\,-}}\left(\frac{1}{e^{\beta E_{1\,+}} - 1}
- \frac{1}{e^{\beta E_{1\,-}} - 1} \right)\right\}
\end{equation}
and
\begin{equation}\label{nonzero modeV}
\left\langle\frac{a_k^{*}a_k}{V}
\right\rangle_{H_{1 \Lambda}}\hskip -0.4 cm ( \mu) =
\frac{1}{V}\frac{1}{e^{\beta (\epsilon _{k}- \mu)} -
1}
\end{equation}
for $k\neq 0,Q$. Thus, since $E_{1\,-} (\mu , \Omega) > 0$ for
$\mu<\mu_c$, we get for the total particle density
\begin{equation}\label{part-dens22}
\lim_{V\rightarrow\infty}\left\langle\frac{{\tilde N}_\Lambda}{V}
\right\rangle_{H_{1 \Lambda}}\hskip -0.4 cm ( \mu) =
\lim_{V\rightarrow\infty}\frac{1}{V}\sum_{k\in \Lambda,\,k \neq 0}\frac{1}{e^{\beta (\epsilon _{k}-
\mu)} - 1}= \rho_0( \mu).
\end{equation}
Therefore as in the PBG we have to fix the finite volume chemical potential
$\mu_V$ by the equation
\begin{equation}\label{part-dens22}
\rho=\left\langle\frac{{\tilde N}_\Lambda}{V}
\right\rangle_{H_{1 \Lambda}}\hskip -0.4 cm( \mu_V).
\end{equation}
and to distinguish two cases, $\rho \leq \rho_{c}$ and $\rho > \rho_{c}$
where
\begin{equation}\label{critdens1}
\rho_{c}\equiv \rho_0(\mu_c ) .
\end{equation}
Notice that by (\ref{PBG-part-dens1}) if $\mu_c<0$, then $\rho_{c}< \infty$ even for $d=1,2$.

\noindent \textbf{(a)} \textit{Case} : $\rho \leq \rho_{c}$. Using
(\ref{part-dens22}) we see that in this case $\mu_V\to \mu_\infty$,
where $\mu_\infty < \mu_c $ is the unique solution of
\begin{equation}\label{ro-crit}
\rho = \rho_0(\mu).
\end{equation}
Since for these values of $\mu$,  $E_{1\,-} (\mu , \Omega) $ is
bounded below away from zero, by (\ref{zero modeV}) we also get for
the expectation of the $Q$ mode occupation particle density
\begin{equation}\label{zero mode}
\lim_{V\rightarrow\infty}\left\langle\frac{a_{Q}^{*}a_{Q}}{V}
\right\rangle_{H_{1 \Lambda}}\hskip -0.4cm ( \mu_V) =  0 .
\end{equation}
Similarly we get
\begin{eqnarray}\label{omega-mode}
&&\lim_{V\rightarrow\infty}\left\langle\frac{b^{*} b}{V}
\right\rangle_{H_{1 \Lambda}}\hskip -0.4cm ( \mu_V) = \\
&&\lim_{V\rightarrow\infty}\frac{1}{2V}\left\{\left(\frac{1}{e^{\beta
E_{1\,+}} - 1} + \frac{1}{e^{\beta E_{1\,-}} - 1}\right) +
\frac{\Omega+\mu -\epsilon_Q}{E_{1\,+} - E_{1\,-}}\left(\frac{1}{e^{\beta
E_{1\,+}} - 1} - \frac{1}{e^{\beta E_{1\,-}} - 1} \right)\right\}
= 0 ,\nonumber
\end{eqnarray}
for the density of photons.

\noindent \textbf{(b)} \textit{Case} : $\rho > \rho_{c}$. As above
the analysis is similar to that for the PBG. Using (\ref{zero modeV}) -
(\ref{part-dens22}) we see that for $V\rightarrow\infty$, $\mu_V$
takes the form:
\begin{equation}\label{mu-V}
\mu_\Lambda= \mu_c - \frac{1}{V}\,\,\frac{1}{\beta (\rho -
\rho_{c})} + {\rm O}(1/V^2) ,
\end{equation}
and
\begin{equation}\label{lim-E-}
E_{1\,-}(\mu_V,\Omega) = \frac{1}{V}\,\,\frac{1}{\beta (\rho -
\rho_{c})}\,\,\frac{\Omega}{ \Omega - \mu_c +\epsilon_Q}+ {\rm O}(1/V^2).
\end{equation}
Then by (\ref{zero modeV}) and (\ref{lim-E-}) we get instead of zero (see \ref{zero mode}):
\begin{equation}\label{BEC}
\lim_{V\rightarrow\infty}\left\langle\frac{a_{Q}^{*}a_{Q}}{V}
\right\rangle_{H_{1 \Lambda}}\hskip -0.4cm (\mu_V) = \rho -\rho_{c},
\end{equation}
implying the occurrence of BEC of the Bose gas in the $Q$-mode. Similarly,
instead of zero as in (\ref{omega-mode}) we obtain
\begin{equation}\label{b-BEC}
\lim_{V\rightarrow\infty}\left\langle\frac{b^{*}b}{V}
\right\rangle_{H_{1 \Lambda}}\hskip -0.4cm  (\mu_V)= \frac{g_1 ^2}{4
\Omega^2}(\rho - \rho_{c})\,,
\end{equation}
that is, one has \textit{condensation of photons} occurring
simultaneously with BEC (\ref{BEC}). This correlation can also be
seen from the limit of the boson-photon correlation
(\textit{entangling}) function:
\begin{eqnarray}\label{corr}
&&\lim_{V\rightarrow\infty}\left\langle\frac{a_{Q}^{*}b}{V}
\right\rangle_{H_{1 \Lambda}}\hskip -0.4cm  (\mu_V) = \nonumber
\\
&&=- \lim_{V\rightarrow\infty} \frac{g_1}{2V(\Omega+\mu_V -\epsilon_Q)}
\left\{\left(e^{\beta E_{1\,+}} - 1\right)^{-1} -
\left(e^{\beta E_{1\,-}}- 1\right)^{-1} \right\} =  \nonumber \\
&&=
\begin{cases}
-g_1\,(\rho - \rho_{c})/2 \Omega &\text{for}\ \rho > \rho_{c}, \\
0 & \text{for}\ \rho \leq \rho_{c}.
\end{cases}
\end{eqnarray}
\\
\\
\noindent \textbf{2.2} Now we condider our second model
$H_{2\,\Lambda}$ We define new boson operators $\xi_{2}^*
, \xi_{2}$ and $\eta_{2}^* , \eta_{2}$ by the canonical Bogoliubov
transformation:
\begin{eqnarray}\label{can-trans-2}
&& \xi_2 = \cosh \phi \,\,\,a_{Q'} - \sinh \phi \,\,\, b \\
&& \eta_2 = \cosh \phi \,\,\,b  - \sinh \phi \,\,\,a_{Q'}
\nonumber \,.
\end{eqnarray}
where $\phi$ satisfies the equation
\begin{equation}\label{equ-theta-2}
    \tanh\,2\phi = - \,\frac{g_2}{\Omega + \epsilon_{Q'} - \mu}.
\end{equation}

In terms of these operators we can write the Hamiltonian
$H_{2\,\Lambda}(\mu)$ in the form
\begin{eqnarray}\label{mu-Ham2-diag}
&&H_{2\,\Lambda}(\mu) = \sum_{k\in \Lambda ^{*},\, k\neq 0,Q'}(\epsilon _{k}- \mu) a_{k}^{*}a_{k} + E_{2\,+} (\mu
, \Omega) \xi_2^* \xi_2 + E_{2\,-} (\mu , \Omega) \eta_2^*
\eta_2\nonumber\\
&& \hskip 7cm+\frac{1}{2}\{(E_{2\,-}+E_{2\,-})-(\epsilon_{Q'}-\mu +\Omega)\}\,\,,
\end{eqnarray}
where
\begin{eqnarray}\label{E2,2}
&& E_{2\,+} (\mu , \Omega) = \frac{1}{2}\left\{(\epsilon_{Q'} - \mu - \Omega) +\sqrt{(\epsilon_{Q'}- \mu +\Omega )^2 -
g_2^2}\right\} \\
&& E_{2\,-} (\mu , \Omega) = \frac{1}{2}\left\{-(\epsilon_{Q'} - \mu - \Omega) +\sqrt{(\epsilon_{Q'}- \mu +\Omega )^2 -
g_2^2}\right\}  \nonumber \,\,.
\end{eqnarray}
The thermodynamic stability of the Hamiltonian $H_{2\,\Lambda}(\mu)$,
always gives
\begin{equation}\label{stab2}
\mu\leq  \mu _c (g_2,\Omega)\,\,
\end{equation}
where
\begin{equation}\label{stab3}
\mu_c(g_2,\Omega)\equiv \min\left\{0\,,\, -\frac{g_2^2}{4\Omega} +
\epsilon_{Q'}\right\}.
\end{equation}
As in (\ref{critdens1}) we let
\begin{equation}\label{critdens2}
\rho_c\equiv\rho( \mu _c ).
\end{equation}
We have for $\mu < \mu _c
$,
\begin{eqnarray}\label{zero modeV2}
\left\langle a_{{Q'}}^{*}a_{{Q'}}
\right\rangle_{H_{2 \Lambda}}\hskip -0.2 cm ( \mu)
=\Bigg\{\left(\frac{1}{e^{\beta E_{2\,+}} - 1} -
\frac{1}{e^{\beta E_{2\,-}} - 1}\right) +\frac{\epsilon_{Q'}-\mu+\Omega}{E_{2\,+} + E_{2\,-}}\left(\frac{1}{e^{\beta E_{2\,+}} - 1}
+\frac{1}{e^{\beta E_{2\,-}} - 1} \right)\nonumber \\
+\frac{\epsilon_{Q'}-\mu+\Omega}{E_{2\,+} + E_{2\,-}}-1\Bigg\},
\end{eqnarray}
\begin{equation}\label{nonzero modeV2}
\left\langle a_k^{*}a_k\right\rangle_{H_{2 \Lambda}}\hskip -0.2 cm ( \mu) =
\frac{1}{e^{\beta (\epsilon _{k}- \mu)} -1}\ \ \ \ \text{for}\ k\neq 0,{Q'},
\end{equation}
\begin{eqnarray}\label{omega-mode2}
&&\left\langle b^{*} b
\right\rangle_{H_{2 \Lambda}}\hskip -0.2cm ( \mu) = \Bigg\{-\left(\frac{1}{e^{\beta E_{2\,+}} - 1} - \frac{1}{e^{\beta E_{2\,-}} - 1}\right) +
\frac{\epsilon_{Q'}-\mu+\Omega}{E_{2\,+} + E_{1\,-}}\left(\frac{1}{e^{\beta
E_{2\,+}} - 1}+ \frac{1}{e^{\beta E_{2\,-}} - 1} \right)\nonumber\\
&&\hskip 11cm+\frac{\epsilon_{Q'}-\mu+\Omega}{E_{2\,+} + E_{1\,-}}-1\Bigg\}
\end{eqnarray}
and
\begin{eqnarray}\label{corr222}
&& \left\langle a_{Q'}^*b^*\right\rangle_{H_{2 \Lambda}}\hskip -0.2cm  (\mu) = \frac{g_2}{2\sqrt{(\epsilon_{Q'}-\mu+\Omega)^2-g_2^2}}
\left\{\left(e^{\beta E_{2\,+}} - 1\right)^{-1} +
\left(e^{\beta E_{2\,-}}- 1\right)^{-1} +1\right\} .
\end{eqnarray}
If $\epsilon_{Q'}\geq g_2^2/4\Omega$ then $\mu_c(g_2,\Omega)=0$ and we return to the PBG with no condensation in the $Q'$-mode.
If $\epsilon_{Q'}< g_2^2/4\Omega$ we have to study two cases $g_2^2\geq 4\Omega^2$ and $g_2^2<4\Omega^2$.
Consider first the case $g_2^2\geq 4\Omega^2$.

\noindent \textbf{(a)} \textit{Case} : $\rho \leq \rho_{c}$\ \
Proceeding in a similar manner as for the first model we get
\begin{equation}\label{zero mode2}
\lim_{V\rightarrow\infty}\left\langle\frac{a_{{Q'}}^{*}a_{{Q'}}}{V}\right\rangle_{H_{2 \Lambda}}\hskip -0.4cm ( \mu_V) =
\lim_{V\rightarrow\infty}\left\langle\frac{b^*b}{V}\right\rangle_{H_{2 \Lambda}}\hskip -0.4cm ( \mu_V) =
\lim_{V\rightarrow\infty}\left\langle\frac{a_{{Q'}}^{*}b^*}{V}
\right\rangle_{H_{2 \Lambda}}\hskip -0.4cm  (\mu_V)=0 .
\end{equation}

\noindent \textbf{(b)} \textit{Case} : $\rho > \rho_{c}$\ \  As for the first case here we obtain
\begin{equation}\label{BEC2}
\lim_{V\rightarrow\infty}\left\langle\frac{a_{{Q'}}^{*}a_{{Q'}}}{V}
\right\rangle_{H_{2 \Lambda}}\hskip -0.4cm (\mu_V) = \rho -\rho_{c},
\end{equation}
\begin{equation}\label{b-BEC2}
\lim_{V\rightarrow\infty}\left\langle\frac{b^{*}b}{V}
\right\rangle_{H_{2 \Lambda}}\hskip -0.4cm  (\mu_V)= \frac{g_2 ^2}{4
\Omega^2}(\rho - \rho_{c})\,,
\end{equation}
and
\begin{eqnarray}\label{corr2}
&&\lim_{V\rightarrow\infty}\left\langle\frac{a_{{Q'}}^{*}b^*}{V}
\right\rangle_{H_{2 \Lambda}}\hskip -0.4cm  (\mu_V) =\frac{g_2}{2\Omega}\,(\rho - \rho_{c}).
\end{eqnarray}\\
\noindent
In the case $g_2^2<4\Omega^2$ we find that the limits have the same form but with $g_2$ and
$\Omega$ interchanged.


\noindent \textbf{3.} \ Clearly in our models the superradiance
(see e.g. \cite{D} and \cite{AEI}) is directly related with BEC as
is explicitly seen for example by comparing the formul{\ae}
(\ref{BEC}) and (\ref{b-BEC}). Hence in spite of the simplicity of
the models (\ref{models 1-2}) they manifest an interesting
cooperative phenomenon. The presence of the interaction between
the Bose gas and radiation, compared to the PBG, occurs in both
models at a lower critical density $\rho_{c}=\rho_0(\epsilon_Q -
g^2/4 \Omega) < \rho_{0 c}=\rho_0( 0)$. Moreover the condensation
in these models takes place not only in dimension $d\geq 3$, but
also in dimensions $d=1$ and $d=2$. This shows clearly that the
presence of radiation enhances the process of condensation in the
Bose gas.

It is also interesting to remark the value of the entangling
boson-photon interaction energy in the two models, which one reads
off from (\ref{corr}) and (\ref{corr2}):
\begin{equation}\label{cond-energy}
\lim_{V\rightarrow\infty}\left\langle\frac{U_{1,2\,\Lambda}}{V}
\right\rangle_{H_{1,2\,\Lambda}}\hskip -0.4 cm (\mu_V) = \mp
\frac{g_{1,2}^2}{2 \Omega}(\rho - \rho_{c}).
\end{equation}
For the first model based on the \textit{minimal coupling} we
obtain the \textit{negative} interaction energy (bound state) in
the presence of condensates, which is well-known
\cite{HL}-\cite{FSV}, \cite{AEI}. Whereas for the second model
this interaction energy is \textit{positive}, which is a
completely different type of entanglement.

These aspects of our results make contact with recent interests in
entangled atom-photon states generated in BEC-superradiance
experiments, see e.g. \cite{K}, addressed to a variety of
applications like tests of Bell inequalities, quantum cryptography
and quantum teleportation.

{\bf Acknowledgements:} Two of the authors (JVP and AV) wish to
thank the Centre Physique Th\'eorique, CNRS-Luminy  for their kind
hospitality while this work was being realized. JVP wishes to
thank University College Dublin for the award of a President's
Research Fellowship.


\end{document}